\documentclass[12pt]{article}

\usepackage{amsfonts}

\author{Justin Malecki\footnote{email: jjmaleck@physics.ubc.ca} \\ \textit{Perimeter
Institute for Theoretical Physics} \\ \textit{Waterloo, Canada}}
\title{Inflationary Quantum Cosmology: General Framework and Exact Bianchi I Solution}
\date{}

\begin{document}

\maketitle

\begin{abstract}
Using the methods of loop quantum
gravity, we derive a framework for describing an inflationary,
homogeneous universe in a purely quantum theory.  The classical
model is formulated in terms of the Ashtekar-Sen connection
variables for a general subclass of Bianchi class A spacetimes.
This formulation then provides a means to develop a corresponding
quantum theory. An exact solution is derived for the classical
Bianchi type I model and the corresponding semi-classical quantum
state is found to also be an exact solution to the quantum theory.
The development of a normalizable quantum state from this exact
solution is presented and the implications for the normalizability
of the Kodama state discussed.  We briefly consider some
consequences of such a quantum cosmological framework and show
that the quantum scale factor of the universe has a continuous
spectrum in this model.  This result may suggest further
modifications that are required to build a more accurate theory.
We close this study by suggesting a variety of directions in which
to take the results presented.
\end{abstract}

\section{Introduction}
\label{sec:intro} The theory of cosmological
inflation~\cite{liddle} has seen continued success since its
introduction, most notably in its accurate description of the
observed fluctuations of the cosmic microwave background (CMB).
Inflation is based on a classical theory of cosmology which is
generally thought to be invalid in regimes close to the big bang
where quantum gravitational dynamics must be considered.  In this paper,
we develop an intrinsically quantum
description of inflation to try and understand the role
of quantum gravity in the early universe.  Our model consists of a
spatially homogeneous universe with scalar field  matter as this provides the
necessary exponential expansion of inflation (with an appropriate choice of scalar
field potential).

A theory of quantum cosmology should be based on a more general theory
of quantum gravity.
Such a theory of quantum gravity that has proved very successful
in recent years is the theory of loop quantum
gravity.\footnote{Thorough, technical introductions to loop quantum
gravity are available~\cite{carlo, thiemann} as well as shorter,
less technical reviews~\cite{carlorev, leerev}, the latter
being closely related to the work discussed in this paper.}
The starting point for the general theory of loop quantum gravity is the
Ashtekar-Sen formulation of general
relativity~\cite{ashvar, ashbook} which greatly simplifies the
dynamical equations and allows for a well defined quantization of
the classical theory. While this quantization procedure is a very
general and powerful construction, we follow a simpler route by
imposing homogeneity of the underlying spacetime.  This allows us
to circumvent some of the regularization procedures that are
necessary in the full theory.

The application of loop quantum gravity to cosmological scenarios
has already seen many intriguing results in the work of Martin Bojowald and others (see~\cite{bojrev} for a review of the
formulation and the main results).  They show that the discreteness of the spectrum of the volume operator leads to many novel effects including a well-defined evolution through the classical big bang singularity~\cite{bojsing}.  In this paper, we use a similar approach to that of Bojowald to build our inflationary model with the advantage that
we are able to find an exact quantum state to describe an
inflationary universe.

The framework for this inflationary quantum cosmological model was
first introduced in~\cite{lqginf}.  In that study, a quantum state
for a flat, isotropic spacetime was derived and discussed.  It was
found to be normalizable and related to a well-known exact quantum
state of the general theory of loop quantum gravity, the Kodama
state~\cite{kodama}. The purpose of the following paper is to present a
generalization of the previous isotropic model to a broader class of homogeneous
spacetimes
and to derive exact quantum states for a subclass of these models.

The starting point for building our model is the classical
Hamiltonian framework of~\cite{lqginf} which is summarized here in
section~\ref{sec:scalarham}.  In section~\ref{sec:classa}, we
specialize this framework to consider only homogeneous universes
and discuss the procedure of quantization.  We then focus on a
single class of homogeneous universe, Bianchi type I, in
section~\ref{sec:bianchi1} where a family of exact classical and
quantum solutions are derived.  It is shown in
section~\ref{sec:normal} that these solutions can form a
normalizable state and we discuss the possible implications for
the Kodama state.  The specialization of these Bianchi I states to flat, isotropic
spacetimes is presented
in~\ref{sec:flatiso} so as to compare with the analogous state
of~\cite{lqginf}. We summarize our results in
section~\ref{sec:disc} and suggest some directions for future
research.  Unless otherwise stated, we work in units in which $c =
\hbar = 1$ so that the Planck length is $\ell_p^2 = G$.

\section{Classical Hamiltonian Scalar Field Model}
\label{sec:scalarham} The formalism of loop quantum gravity
traditionally begins by describing general relativity as a
constrained Hamiltonian theory in terms of Ashtekar-Sen
variables~\cite{ashvar, ashbook} before applying the quantization
procedure of Dirac~\cite{dirac}.  Since we are interested in
modelling inflation by including a scalar field into our model, it
is possible to recast this classical framework into a more
familiar Hamiltonian language by using the additional matter
degrees of freedom.  This derivation was first presented
in~\cite{lqginf} and so we briefly summarize the procedure here.

The phase space variables used to describe the geometry of a
spatial slice of spacetime are a complex, self-dual $SU(2)$
connection $A_a^i$ and its conjugate momentum $E_i^a$.  Here and
throughout we use indices $a, b, c, \ldots \in \{ 1, 2, 3 \}$ as
spatial indices and $i, j, k, \ldots \in \{ 1, 2, 3 \}$ as $su(2)$
indices, the latter being raised and lowered by the Kronecker
delta $\delta_{ij}$. The momentum $E_i^a$ is the densitized
spatial triad
\begin{equation}
E_i^a = \det(e_b^j) (e^{-1})_i^a
\end{equation}
where  $e_a^i$ is the triad or frame field on the spatial slice
and $(e^{-1})_i^a$ its inverse. The connection can be expressed in
terms of the three dimensional spin connection $\Gamma_a^i$ of the
frame field and the extrinsic curvature $K_a^b$ of the spatial
slice by
\begin{equation}
\label{geoa} A_a^i = \Gamma_a^i + i e_b^i K_a^b.
\end{equation}
The canonical relationship between these phase space variables is
given by the Poisson bracket
\begin{equation}
\label{pois1} \{ A_a^i(x), \ E_j^b(y) \} = i \ell_p^2 \delta_a^b
\delta_j^i \delta^3(x, y)
\end{equation}
where $x$ and $y$ are coordinates on the spatial slice.  Since the
triad and spin connection are both real quantities, it may be
necessary to impose the reality conditions
\begin{eqnarray}
\label{reality1} A_a^i + \bar{A_a^i} & = & 2 \Gamma_a^i \\
\label{reality2} E_i^a - \bar{E}_i^a & = & 0
\end{eqnarray}
on the phase space variables where $\bar{\phantom{A_a^i}}$ denotes
complex conjugation.

As in conventional inflationary cosmology, we consider gravity
coupled to a scalar field $\phi$ in a potential $V(\phi)$ with
conjugate scalar momentum $\pi$.  The gauge, diffeomorphism, and
Hamiltonian constraints can be written respectively as
\begin{eqnarray}
\label{gaugecon} G_i & = & D_a E^a_i =  \partial_a E^a_i +
\epsilon_{ij}^{\phantom{ij}k} A_a^j E^a_k = 0, \\
\label{diffcon} H_a & = & E^b_i F_{ab}^i + \pi \partial_a \phi = 0,
\end{eqnarray}
and
\begin{equation}
\label{hamcon} \mathcal{H} = \frac{1}{\ell_p^2} \epsilon^{ijk}
E^a_i E^b_j \left( F_{abk} + \frac{\ell_p^2 V(\phi)}{3}
\epsilon_{abc}E^c_k \right) + \frac{1}{2} \pi^2 + \frac{1}{2}
E^{ai} E^b_i (\partial_a \phi) (\partial_b \phi) = 0
\end{equation}
where $\epsilon_{ijk}$ is the totally antisymmetric, Levi-Civita tensor, $F_{abk}$
is the curvature of the connection $A_a^i$, and the cosmological
constant $\Lambda$ has been absorbed into the potential $V(\phi)$.

We choose a decomposition of the spacetime manifold $\mathcal{M}$
into $\mathbb{R} \times \mathcal{S}$ by choosing space
$\mathcal{S}$ to be defined as a constant $\phi$ hypersurface. The
limitations of such a gauge choice are discussed in~\cite{lqginf}
but are not important for the results of this paper.  It is
sufficient to know that such a gauge choice is valid during the
inflationary phase but may break down after the exponential
expansion.

To implement this gauge, we choose a basis of spatial vector
fields $\{ \partial_a \}_{a = 1}^3$ such that
\begin{equation}
\label{phistrict}
\partial_a \phi = 0
\end{equation}
and demand that this constraint be preserved in time by demanding
\begin{equation}
\label{tpres} \{ \partial_a \phi, \ \mathcal{H}(\tilde{N}) \} =
\partial_a(\tilde{N} \pi) = 0
\end{equation}
where $\mathcal{H}(\tilde{N}) = \int_\mathcal{S} \tilde{N}
\mathcal{H}$ is the integrated Hamiltonian constraint that
generates diffeomorphisms in the time direction.
Equation~(\ref{tpres}) implies that the \emph{densitized} lapse
$\tilde{N}$ must take the form
\begin{equation}
\tilde{N} = \frac{k}{\pi}
\end{equation}
where $k$ is a constant, chosen to be $k = \ell_p^{-2}$ so that
the lapse is dimensionless.

The diffeomorphism and gauge constraints~(\ref{diffcon})
and~(\ref{gaugecon}) will be solved automatically when we
eventually reduce to the homogeneous models and so we focus only
on satisfying the Hamiltonian constraint~(\ref{hamcon}).  In light
of the time gauge condition~(\ref{phistrict}) we can solve the
Hamiltonian constraint for $\pi$ to get
\begin{equation}
\label{pi} \pi = \pm \sqrt{- \frac{2}{\ell_p^2} \epsilon^{ijk} E^a_i
E^b_j \left( F_{abk} + \frac{\ell_p^2 V(\phi)}{3}
\epsilon_{abc}E^c_k \right)}.
\end{equation}

Combining~(\ref{hamcon}), (\ref{phistrict}), and~(\ref{pi}) one
can show that the integrated Hamiltonian constraint is
\begin{equation}
\label{hammer} \mathcal{H}(\tilde{N}) = \frac{1}{2 \ell_p^2}
\int_\mathcal{S} \pi \mp \frac{1}{\sqrt{2} \ell_p^2}
\int_\mathcal{S} \sqrt{- \frac{1}{\ell_p^2} \epsilon^{ijk} E^a_i
E^b_j \left( F_{abk} + \frac{\ell_p^2 V(\phi)}{3}
\epsilon_{abc}E^c_k \right)}.
\end{equation}
We use $\phi$ to define a time coordinate\footnote{Note that
slow-roll inflation~\protect\cite{liddle} is achieved by a
\emph{decreasing} $\phi$ (\textit{i.e.} $\phi$ is ``rolling down
the hill'' of the potential).  This means that the \emph{forward}
progression of time corresponds to a \emph{decreasing} $T$.}
\begin{equation}
\label{time} T := \ell_p^2 \phi
\end{equation}
with conjugate momentum
\begin{equation}
\label{tmom} P  := \frac{1}{l_p^2} \int_\mathcal{S} \pi
\end{equation}
\begin{equation}
\label{tpois} \Rightarrow  \{T, \ P \} = 1
\end{equation}
so that setting the integrated constraint $\mathcal{H}(\tilde{N})
= 0$ in~(\ref{hammer}) implies
\begin{equation}
\label{hamrel} H = P
\end{equation}
where we recognize
\begin{equation}
\label{gaugeham} H := \pm \frac{\sqrt{2}}{\ell_p^2} \int_\mathcal{S}
\sqrt{-\frac{1}{\ell_p^2} \epsilon^{ijk} E^a_i E^b_j \left( F_{abk}
+ \frac{\ell_p^2 V(T)}{3} \epsilon_{abc}E^c_k \right)}
\end{equation}
as the Hamiltonian that generates evolution in the parameter $T$.
Herein we will use $T$ instead of $\phi$ and denote the scalar
field potential as $V(T)$ to mean the original scalar field
potential $V(\phi)$ evaluated at $\phi = T/\ell_p^2$.

To summarize, by choosing an appropriate coordinate gauge, we have
been able transform the constrained Hamiltonian theory of general
relativity into a theory whose dynamics are generated by a
traditional Hamiltonian functional~(\ref{gaugeham}).  This will be
the starting point for reducing to homogeneous spacetimes and,
eventually, quantizing to obtain a quantum theory for a
homogeneous universe.

\section{Bianchi Class A Models}
\label{sec:classa} We now proceed to specialize the Hamiltonian
formalism of the previous section to consider only the so-called
Bianchi class A models. The result will be a simpler theory that
can immediately be quantized, providing a quantum framework to
describe a homogeneous, inflationary universe.

A spatially homogeneous spacetime\footnote{A complete description
of homogeneous spacetimes is given in~\cite{ryshep} where the
reader is referred to for a more detail derivation of the facts
stated here.} is characterized by a 3-dimensional symmetry group
acting transitively on the spatial manifold $\mathcal{S}$. Such a
group can be described by its algebra with structure constants
$c^I_{\phantom{I}JK}$ which can, in general, be written as
\begin{equation}
c^K_{\phantom{K}IJ} = M^{KL} \epsilon_{LIJ} + \delta^K_{[I} A_{J]}
\end{equation}
where $M^{KL}$ and $A_J$ are any tensors satisfying
\begin{equation}
M^{IJ} A_J = 0.
\end{equation}
The Bianchi class A models are those in which $A_J = 0$.  This
class consists of 6 models (up to isomorphism) and we further
restrict our consideration to those 5 whose structure constants
can be written in the form
\begin{equation}
\label{diagconst} c^I_{\phantom{I}JK} = n^{(I)}
\epsilon^{I}_{\phantom{I}JK}
\end{equation}
where the brackets around an index indicate that it is not to be
summed over.  Herein, we will use the term Bianchi class A to mean
this subset of 5 models.

It can be shown~\cite{ryshep} that one can always construct a
basis of 1-form fields $\{ \omega^I \}$ with dual vector fields
$\{ X_I \}$ which are left-invariant under the action of the
symmetry group (\textit{i.e.} a basis that does not change when
acted upon by a group element from the left).  Such a basis is
uniquely determined by demanding that it satisfy the ``curl''
relations
\begin{equation}
\label{domegaabs} d \omega^I = - \frac{1}{2} c^I_{\phantom{I}JK}
\omega^J \wedge \omega^K
\end{equation}
and that its Lie derivative along timelike vector fields
orthogonal to the spatial slices vanish.  It will prove
advantageous to utilize such an invariant basis in order to
simplify the calculations.

Following the procedure described in~\cite{kodbianchi, homobojo},
we expand the $A$ and $E$ fields in terms of these left-invariant
bases. The connection and momentum can then be written as
\begin{equation}
\label{homovar} A_a^i(x, T) = c_{(I)}(T) \Lambda^i_I \omega^I_a,
\qquad E_i^a(x, T) = p^{(I)}(T) \Lambda^I_i X^a_I
\end{equation}
where $\Lambda \in SO(3)$ and the functions $\{c_I\}$ and
$\{p^I\}$ are constant on a given spatial slice and only depend on
the time variable~(\ref{time}). Please note that~(\ref{homovar})
is not the most general form of the homogeneous variables but is,
in fact, a diagonalized form that is well suited to the study of
Bianchi class A models. The $SO(3)$ matrices are, essentially, a
rotation of the invariant basis introduced in order to obtain this
diagonal form of the connection and momentum. These matrices are a
statement of the gauge degrees of freedom in the $A$ and $E$
variables. For the computations that are to follow, we will often
use the following properties of $SO(3)$ matrices:
\begin{equation}
\label{so3tricks} \Lambda^i_I \Lambda^J_i = \delta_I^J, \qquad
\epsilon_{ijk}\Lambda^i_I \Lambda^j_J \Lambda^k_K =
\epsilon_{IJK}.
\end{equation}

In order to avoid infinite quantities arising from the
(possible) infinitude of
space, we consider a compact submanifold $\Sigma \subset
\mathcal{S}$ of the full spatial manifold $\mathcal{S}$. Since the
spatial manifold is, by definition, homogeneous, no generality is
lost by this reduction. We parametrize this submanifold choice by
defining a dimensionless parameter, $R$, such that
\begin{equation}
R^3 := \frac{1}{\ell_p^3} \int_\Sigma
\end{equation}
where the volume element on $\Sigma$ is understood. The
$\ell_p^{-3}$ normalization is simply an arbitrary length scale,
chosen to be the Planck length.  $R$ can then be considered a
parameter of the model that measures the size of our sample of
space.

The reduction~(\ref{homovar}) of the full theory to Bianchi class
A models reduces general relativity from a Hamiltonian theory with
an infinite number of degrees of freedom to one with only three
degrees of freedom, $c_1$, $c_2$, and $c_3$ and their conjugate
momenta $p^1$, $p^2$, and $p^3$.  To determine their canonical
relationship, we must first define $c_I$ and $p^I$ in terms of the
variables $A_a^i$ and $E_i^a$.  A convenient definition is
\begin{eqnarray}
c_I & = & \frac{1}{\ell_p^3 R^3} \int_\Sigma A_a^i \Lambda_i^{(I)}
X_I^a \\
p^I & = & \frac{1}{\ell_p^3 R^3} \int_\Sigma E_i^a \Lambda_{(I)}^i
\omega^I_a.
\end{eqnarray}
With these definitions we can use the relation~(\ref{pois1}) to
compute the Poisson bracket
\begin{equation}
\label{pois} \{ c_I(T), \ \frac{R^3 \ell_p}{i}p^J(T) \} =
\delta_I^J.
\end{equation}
Hence, we recognize $(R^3 \ell_p / i) p^I(T)$ as being the true
canonical momenta to the variables $c_I(T)$.  Of course, the
Poisson bracket between different $c_I$ variables and between
different $p^J$ variables respectively vanishes.

In order to write the diffeomorphism and Hamiltonian constraints
in terms of the new phase space variables $(c_I, p^J)$, we must
first calculate the curvature of the homogeneous connection
\begin{equation}
F = dA + \frac{1}{2} [A, A].
\end{equation}
By expanding $A = A^i \tau_i$ in terms of the $su(2)$ basis $\tau_i := -\frac{i}{2} \sigma_i$ (where $\sigma_i$ are the Pauli matrices) and using
the identities~(\ref{domegaabs}) and~(\ref{so3tricks}) we can work
out the curvature components in the left-invariant basis to be
\begin{equation}
\label{curvcoord} F^i_{JL} = -n^{(I)} c_{(I)}
\epsilon^I_{\phantom{I}JL} \Lambda_I^i + c_{(J)} c_{(L)}
\epsilon^i_{\phantom{i}jk} \Lambda_J^j \Lambda_L^k.
\end{equation}
Here, the constants $\{ n^I \}_{I = 1}^3$ are the structure
constants of the symmetry group defined in~(\ref{diagconst}).

We can now use these curvature components and the homogeneous
momentum field~(\ref{homovar}) to calculate the diffeomorphism
constraint~(\ref{diffcon}) in the left invariant basis.  The
result is the expression
\begin{equation}
H_I = p^{(J)} \Lambda^J_i \left( -n^{(I)} c_{(I)}
\epsilon^L_{\phantom{L}IJ} \Lambda_I^L + c_{(I)} c_{(J)}
\epsilon^i_{\phantom{i}jk} \Lambda_I^j \Lambda_J^k \right)
\end{equation}
which, by applying the first relation in~(\ref{so3tricks}), can be
shown to vanish identically for all Bianchi class A models. So, by
reducing our field variables to an intrinsically homogeneous and
diagonalized form, we have also, at the same time, automatically
satisfied the diffeomorphism constraint.  Note that this result is
not necessarily true for more general forms of the homogeneous
variables.

An expression for the Hamiltonian~(\ref{gaugeham}) in terms of the
new phase space variables can be obtained in a similar fashion.
That is, by substituting the curvature~(\ref{curvcoord}) and the
components of the homogeneous variables~(\ref{homovar}) into the
Hamiltonian and applying the $SO(3)$ relations~(\ref{so3tricks})
one obtains the Hamiltonian
\begin{equation}
\label{hamiltonian} H(c_I, p^J; T) = \pm 2 R^3 \left[ p^1 p^2 c_1
c_2 \left( n^3 \frac{c_3}{c_1 c_2} - 1 - \frac{\ell_p^2 V(T)}{3}
\frac{p^3}{c_1 c_2} \right) + \begin{array}{c}
  \textrm{cyc.} \\
  \textrm{perm.} \\
\end{array}
\right]^{\frac{1}{2}}.
\end{equation}
Recall that $V(T)$ is the potential energy function for the scalar
field $\phi = \ell_p^{-2} T$, kept as an arbitrary function
throughout the paper.

In order to determine the reality condition~(\ref{reality1}) for
the new phase space variables, one must, first, expand the three
dimensional spin connection in terms of the ($SO(3)$ rotated) left
invariant basis
\begin{equation}
\Gamma_a^i = \Gamma_{(I)} \Lambda_I^i \omega^I_a
\end{equation}
where $\Gamma_I$ can be computed~\cite{bojbianc1} to be
\begin{equation}
\label{spincon} \Gamma_I = \frac{1}{2} \left( n^J \frac{p^K}{p^J}
+ n^K \frac{p^J}{p^K} - n^I \frac{p^J p^K}{(p^I)^2} \right) \qquad
\textrm{(no summation on indices)}
\end{equation}
where $(I, J, K)$ are even permutations of (1, 2, 3).  The
resulting reality conditions are then
\begin{eqnarray}
\label{realhomo1} c_I + \overline{c_I} & = & 2 \Gamma_I \\
\label{realhomo2} p_I - \overline{p_I} & = & 0.
\end{eqnarray}
Due to the presence of the $n^I$ parameters in~(\ref{spincon}), we
see that the reality conditions are dependent on which Bianchi
model is chosen.

Our classical Hamiltonian theory for a homogeneous universe with a
scalar field is now completely defined.  We have phase space
variables $c_I$ and $p^I$ with the symplectic
structure~(\ref{pois}), the reality conditions~(\ref{realhomo1})
and~(\ref{realhomo2}) and a time dependent
Hamiltonian~(\ref{hamiltonian}) that governs the dynamics.  The
definitions~(\ref{homovar}) then relate the phase space variables
back to the Ashtekar-Sen variables from which we can then infer
all of the information concerning the spacetime geometry. Since
there are only a finite number of degrees of freedom and no
constraints needing to be satisfied, one can, in principle,
quantize this model in a straightforward manner following the
traditional algorithm of quantizing classical Hamiltonian
systems~\cite{diracprinc}.

\subsection{Zero-Energy Classical Solution}
Before looking at the quantum theory, it would be very desirable
to obtain a general solution to this classical Hamiltonian theory.
As was seen in the calculation of the quantum solution to the
flat, isotropic inflationary model in~\cite{lqginf}, such a
classical solution can be crucial in finding a solution to the
corresponding quantum theory.  This will again be the case in the
following section.

A first attempt at a general, classical solution may be made via
Hamilton-Jacobi theory~\cite{goldstein} in which we seek a
Hamilton-Jacobi function $S(c_1, c_2, c_3; T)$ such that
\begin{equation}
\label{hjrel} p^I = \frac{i}{R^3 \ell_p} \frac{\partial
S}{\partial c_I}, \qquad P = \frac{\partial S}{\partial T}
\end{equation}
where the constant coefficient in the first relation comes from
the non-standard Poisson bracket~(\ref{pois}).  Such a function is
determined by equation~(\ref{hamrel}) which now reads as a partial
differential equation
\begin{eqnarray}
\nonumber \frac{\partial S}{\partial T} & = & H \left( c_I,
\frac{i}{R^3
\ell_p} \frac{\partial S}{\partial c_I}; T \right) \\
\label{genhjeqn} \Rightarrow \frac{\partial S}{\partial T} & = &
\frac{2 i}{\ell_p} \left[ \frac{\partial S}{\partial c_1}
\frac{\partial S}{\partial c_2} c_1 c_2 \left( n^3 \frac{c_3}{c_1
c_2} - 1 - \frac{i \ell_p V(T)}{3 R^3 c_1 c_2} \frac{\partial
S}{\partial c_3} \right) + \begin{array}{c}
  \textrm{cyc.} \\
  \textrm{perm.} \\
\end{array}
\right]^{\frac{1}{2}}
\end{eqnarray}
where we have chosen the positive root of the Hamiltonian.
The determination of such a Hamilton-Jacobi function is equivalent
to solving the equations of motion.

While we are not able to find a completely general solution
to~(\ref{genhjeqn}) for arbitrary values of $n^I$, we are able to
find a ``zero-energy'' function $S_0(c_1, c_2, c_3; T)$, related
to the phase space variables as in~(\ref{hjrel}), such that $H =
0$.  The determining differential equation is the vanishing of the
right hand side of~(\ref{genhjeqn}) which does indeed vanish if
\begin{equation}
\frac{\partial S_0}{\partial c_K} = \frac{3 i R^3}{\ell_p V}
\left( c_I c_J - n^{(K)} c_K  \right)
\end{equation}
where $(I, J, K)$ are cyclic permutations of $(1, 2, 3)$.  Such a
zero-energy function is given as
\begin{equation}
\label{zeroe} S_0(c_1, c_2, c_3; T) = \frac{3 i R^3}{\ell_p V}
\left( c_1 c_2 c_3 - \frac{1}{2} n^I c_I^2 \right)
\end{equation}
plus an irrelevant constant. It is clear that this function does
\emph{not} solve the full Hamilton-Jacobi
equation~(\ref{genhjeqn}) since the time derivative of $S_0$ does
not vanish for arbitrary potentials $V$.  Of course this
\emph{would} be the general solution for the case where $V$ is
just the cosmological constant with no time dependence.

As we will see in the next section (and as was seen
in~\cite{lqginf}), such zero-energy functions are often a useful
basis for an ansatz that can be used to find a general solution to
the Hamilton-Jacobi equation.  While such a strategy proved useful
for the Bianch type I model considered in the next section, it is
not clear how to successfully proceed with such a strategy for
general Bianchi class A models.

\section{Bianchi Type I Classical and Quantum Solutions}
\label{sec:bianchi1} We now turn our attention to the simplest,
specific class A model: Bianchi I.  The symmetry group of Bianchi
I is the three dimensional translation group which is Abelian so
the structure constants vanish and we have $n^I \equiv 0$.
Furthermore, because of this commutativity, we may use the
coordinate 1-forms $dx^I$ as our left-invariant 1-forms,
\begin{equation}
\omega^I = dx^I,
\end{equation}
which clearly satisfy the curl relations~(\ref{domegaabs}).

Since the structure constants vanish for the model under
consideration, it is clear that the 3 dimensional spin
connection~(\ref{spincon}) also vanishes and so the reality
conditions on the configuration space variables read
\begin{equation}
c_I = - \overline{c_I}
\end{equation}
which can trivially be satisfied by defining new variables
\begin{equation}
\label{newc} c^{\textrm{new}}_I := - i c^{\textrm{old}}_I
\end{equation}
where $c^{\textrm{new}}_I$ can now be taken to be purely real
(\textit{i.e.} $c^{\textrm{old}}_I$ are purely imaginary in
Bianchi I).  Herein, we will use, simply, $c_I$ to denote the new,
purely real variables.  Of course, given the trivial reality
conditions~(\ref{realhomo2}) on $p^I$, the conjugate momenta can
be taken to be purely real.  With these new $c_I$ variables, the
Poisson bracket relation~(\ref{pois}) between the phase space
variables now reads
\begin{equation}
\label{realpois} \{ c_I, \ p^J \} = \frac{1}{R^3 \ell_p}
\delta_I^J.
\end{equation}

We can now adapt the derivation of the previous section to this
specific model. Using the real variables and $n^I \equiv 0$ in the
Hamiltonian~(\ref{hamiltonian}) we get the Bianchi I Hamiltonian
\begin{equation}
\label{hamb} H(c_I, p_J; T) = 2 R^3 \left[ p^1 p^2 c_1 c_2 \left(
1 - \frac{\ell_p^2 V(T)}{3} \frac{p^3}{c_1 c_2} \right) +
\begin{array}{c}
  \textrm{cyc.} \\
  \textrm{perm.} \\
\end{array}
\right]^{\frac{1}{2}}
\end{equation}
where, again, we have chosen the positive root.
Given the slightly different Poisson bracket~(\ref{realpois}) for
the real variables, a Hamilton-Jacobi function $S$ is now defined
as
\begin{equation}
\label{hjrelreal} p^I = \frac{1}{R^3 \ell_p} \frac{\partial
S}{\partial c_I}, \qquad P = \frac{\partial S}{\partial T}
\end{equation}
and the corresponding Hamilton-Jacobi equation $P = H$ reads
\begin{equation}
\label{realhjeqn} \frac{\partial S}{\partial T} = \frac{2}{\ell_p}
\left[ \frac{\partial S}{\partial c_1} \frac{\partial S}{\partial
c_2} c_1 c_2 \left( 1 - \frac{\ell_p V(T)}{3 R^3 c_1 c_2}
\frac{\partial S}{\partial c_3} \right) +
\begin{array}{c}
  \textrm{cyc.} \\
  \textrm{perm.} \\
\end{array}
\right]^{\frac{1}{2}}.
\end{equation}

To obtain a classical solution for this model, we begin with the
zero-energy solution~(\ref{zeroe})
\begin{equation}
S_0 = \frac{3 R^3}{\ell_p V} c_1 c_2 c_3
\end{equation}
and take the ansatz
\begin{equation}
\label{su} S_u := \frac{3 R^3}{\ell_p V} c_1 c_2 c_3 (1 + u(T))
\end{equation}
where $u(T)$ is a complex function of T alone.  Upon substituting
this ansatz into the Hamilton-Jacobi equation~(\ref{realhjeqn}) we
find that the $c_I$ variables cancel out on both sides and we are
left with an ordinary differential equation in $u$
\begin{equation}
\label{ueq} \dot{u} = \frac{\dot{V}}{V}(1 + u) + \frac{2i}{\ell_p}
(1 + u) \sqrt{3u}
\end{equation}
where the dot denotes differentiation with respect to $T$. Since
this is an ordinary, differential equation, there exists a one
(complex) dimensional solution space and, therefore, we have
obtained a one-parameter family of solutions to the classical
Bianchi I model via the Hamilton-Jacobi function~(\ref{su}).

Armed with a solution to the classical model we can now proceed to
build a quantum Bianchi I framework following the usual
quantization procedure. Since $c_I$ and $p^I$ are real valued, we
take the corresponding operators $\hat{c}_I$ and $\hat{p}^J$ to be
Hermitian
\begin{equation}
(\hat{c}_I)^\dag = \hat{c}_I, \quad (\hat{p}^I)^\dag = \hat{p}^I.
\end{equation}
The canonical commutation relations between these operators follow
from the classical Poisson brackets and are given as
\begin{eqnarray}
[ \hat{c}_I, \ \hat{p}^J ] & = & i \{ c_I, \ p^J \} = \frac{i}{R^3
\ell_p} \delta_I^J, \\
\left[ \hat{c}_I, \ \hat{c}_J \right] & = & i \{ c_I, \ c^J \} = 0, \\
\left[ \hat{p}^I, \ \hat{p}^J \right] & = & i \{ p_I, \ p^J \} =
0.
\end{eqnarray}
We can represent the vectors in the Hilbert space as functions
$\Psi : \mathbb{R}^3 \to \mathbb{C}$ of the three configuration
space variables $c_1$, $c_2$, and $c_3$.  In this representation,
the phase space operators act as
\begin{equation}
\label{oprep} \hat{c}_I \Psi = c_I \Psi, \qquad \hat{p}^I \Psi = -
\frac{i}{R^3 \ell_p} \frac{\partial}{\partial c_I} \Psi.
\end{equation}
It is easy to check that these satisfy the above commutation
relations.

The next step in this procedure is to construct a Hamiltonian
operator corresponding to the classical expression~(\ref{hamb}).
In terms of the classical variables, we may order the terms of the
Hamiltonian~(\ref{hamb}) as
\begin{equation}
\label{hamord} H = 2 R^3 c_1 c_2 c_3 \left[ \frac{1}{c_2 (c_3)^2}
p^1 \frac{1}{c_1} p^2 \left( 1 - \frac{\ell_p^2 V}{3 c_1 c_2} p^3
\right) +
\begin{array}{c}
  \textrm{cyc.} \\
  \textrm{perm.} \\
\end{array}
\right]^{\frac{1}{2}}.
\end{equation}
Using the operator representation~(\ref{oprep}) we can construct
the corresponding operator as
\begin{equation}
\label{hamop} \hat{H} = - \frac{2 i}{\ell_p} c_1 c_2 c_3 \left[
\frac{1}{c_2 (c_3)^2} \frac{\partial}{\partial c_1} \frac{1}{c_1}
\frac{\partial}{\partial c_2} \left( 1 + \frac{i \ell_p V}{3 R^3
c_1 c_2} \frac{\partial}{\partial c_3} \right) +
\begin{array}{c}
  \textrm{cyc.} \\
  \textrm{perm.} \\
\end{array}
\right]^{\frac{1}{2}}.
\end{equation}
Our goal now is to find a solution $\Psi(c_1, c_2, c_3; T)$ to the
resulting Schr\"odinger equation
\begin{equation}
\label{schro} i \frac{\partial \Psi}{\partial T} = \hat{H} \Psi.
\end{equation}

We introduce the semi-classical state
\begin{equation}
\label{psiu} \Psi_u := e^{i S_u} = e^{\frac{3iR^3}{\ell_p V}c_1
c_2 c_3 (1 + u)}
\end{equation}
where $S_u$ is the classical Hamilton-Jacobi solution~(\ref{su}).
If we denote the operator under the square root of~(\ref{hamop})
as $\hat{\Omega}$ so that
\begin{equation}
\label{hamshort} \hat{H} = -\frac{2i}{\ell_p} c_1 c_2 c_3
\sqrt{\hat{\Omega}}
\end{equation}
then it can be shown by direct computation that $\Psi_u$ is an
eigenvector of $\hat{\Omega}$ with time dependent eigenvalue:
\begin{eqnarray}
\hat{\Omega} \Psi_u & = & 3 u \left( \frac{3 R^3}{\ell_p V} (1 +
u)
\right)^2 \Psi_u \\
\Rightarrow \sqrt{\hat{\Omega}} \Psi_u & = & \sqrt{3 u \left(
\frac{3 R^3}{\ell_p V} (1 + u) \right)^2} \Psi_u.
\end{eqnarray}
Hence, the corresponding action of the Hamiltonian
operator~(\ref{hamshort}) can immediately be determined as
\begin{equation}
\label{hamact} \hat{H} \Psi_u = - \frac{6 R^3 i}{\ell_p^2 V} c_1
c_2 c_3 (1+u) \sqrt{3 u} \Psi_u.
\end{equation}

The surprising fact is that the semi-classical state~(\ref{psiu})
is, in fact, a full solution to the Schr\"odinger
equation~(\ref{schro}). To see this, we compute the time
derivative of $\Psi_u$ to be
\begin{equation}
\frac{\partial \Psi_u}{\partial T} = i \frac{3 R^3}{\ell_p V} c_1
c_2 c_3 \left( - \frac{\dot{V}}{V} (1 + u) + \dot{u} \right)
\Psi_u
\end{equation}
and, upon substitution of the differential equation~(\ref{ueq}),
this becomes
\begin{equation}
\frac{\partial \Psi_u}{\partial T} = - \frac{6 R^3}{\ell_p^2 V}
c_1 c_2 c_3 (1 + u) \sqrt{3 u} \Psi_u.
\end{equation}
Comparison of this with the action of the Hamiltonian
operator~(\ref{hamact}) clearly shows that $\Psi_u$ solves
Schr\"odinger's equation~(\ref{schro}).  Hence,
equation~(\ref{psiu}) gives a one parameter family of quantum states for an
inflationary universe.

\subsection{Normalizability and the Relationship with the Kodama
State} \label{sec:normal} A famous solution to the quantum
constraint equations of quantum gravity with a cosmological
constant is the so-called Kodama state~\cite{kodama}.  For a
general (\textit{i.e.} non-homogeneous) connection, it can be
written as
\begin{equation}
\label{kodama} \Psi_{K}[A] := \mathcal{N} e^{-i S_{CS}[A]} =
\mathcal{N} e^{\frac{3}{2 \Lambda \ell_p^2} \int_\mathcal{S}
Y_{CS}[A]}
\end{equation}
where $Y_{CS}[A]$ is the Chern-Simons invariant
\begin{equation}
\label{chernsimons} Y_{CS}[A] := \frac{1}{2} \textrm{Tr} \left(A
\wedge dA + \frac{2}{3} A \wedge A \wedge A \right)
\end{equation}
and $\mathcal{N}$ is a normalization factor depending only on the
topology of the manifold~\cite{soo}. Similar to the homogeneous,
inflationary solution found above, the Kodama state is also a
semi-classical state corresponding to the Hamilton-Jacobi
functional
\begin{equation}
\label{hjchern} S_{CS}[A] = \frac{3i}{2 \Lambda G}
\int_\mathcal{S} Y_{CS}[A]
\end{equation}
which solves the classical constraint equations of gravity.

One of the motivations for studying our inflationary model was to
try to find states similar to the Kodama state~(\ref{kodama}) in a
framework in which the state's properties could more easily be
calculated.  In this section, we investigate the similarities
between the Kodama state~(\ref{kodama}) and our inflationary
state~(\ref{psiu}) and discuss the relevant properties of the
latter.

In order to see what the Kodama state looks like in a Bianchi
spacetime, we need to compute the Chern-Simons
invariant~(\ref{chernsimons}) using the homogeneous
connection~(\ref{homovar}).  We do this by explicitly expanding
the connection $A$ as $A^i \tau_i$ (recall that $\tau_i = -\frac{i}{2} \sigma_i$) to obtain
\begin{equation}
Y_{CS} = \frac{1}{4}(n^i c_i^2 - 2 c_1 c_2 c_3) \omega^1 \wedge \omega^2 \wedge
\omega^3
\end{equation}
where $\omega^I$ are the left-invariant basis 1-forms.
Substituting this into the Hamilton-Jacobi
functional~(\ref{hjchern}) we get
\begin{equation}
\label{hjchernhomo} S_{CS} = \frac{3 i R^3 \ell_p}{4 \Lambda}
 \left( \frac{1}{2} n^i c_i^2 - c_1 c_2 c_3 \right)
\end{equation}
which is the same as the zero-energy Hamilton-Jacobi function
$S_0$~(\ref{zeroe}) with the scalar potential $V(T)$ chosen to be
constant. Of course, this similarity is not
so surprising. The $S_{CS}$ function was chosen so that the
Hamiltonian constraint vanished while $S_0$ is defined as that
function which causes the Hamiltonian~(\ref{hamiltonian}), which
is the integral of the square root of the Hamiltonian constraint,
to vanish.

Using this result we can specialize the Kodama state to Bianchi I
to obtain
\begin{equation}
\label{kodhomo} \Psi_K = \mathcal{N} e^{-i S_{CS}} = \mathcal{N}
e^{i \frac{3 R^3 \ell_p}{4 \Lambda} c_1 c_2 c_3}.
\end{equation}
Comparing this with the inflationary state $\Psi_u$ from
equation~(\ref{psiu}) we see they are the same state up to the
time varying function $u(T)$ (with the appropriate choice of
constant potential $V(T)$). We also emphasize, once again, that
both the Kodama state $\Psi_K$ and our new state $\Psi_u$ are
semi-classical states as well as full solutions to their
respective theories (the former solves the constraint equations
while the latter solves Schr\"odinger's equation). These
similarities suggest that $\Psi_u$ is a \emph{modified} Kodama
state; it is the analogue of the Kodama state when one couples a
scalar field to the quantum gravitational field.

Due to the fact that the configuration space variables $c_I$ are
real, we may take the inner product between two states $| \Psi
\rangle$ and $| \Phi \rangle$ to be
\begin{equation}
\langle \Psi | \Phi \rangle = \int_{\mathbb{R}^3} dc_1 dc_2 dc_3 \
\overline{\Psi}(c_1, c_2, c_3) \Phi(c_1, c_2, c_3)
\end{equation}
where, as usual, $\Psi(c_1, c_2, c_3) := \langle c_1, c_2, c_3 |
\Psi \rangle$ and $| c_1, c_2, c_3 \rangle$ are the simultaneous
eigenvectors of the $\hat{c}_I$ operators.  Using this inner
product, it is clear that both the Kodama state~(\ref{kodhomo})
and the modified Kodama state~(\ref{psiu}) are both continuously
normalizable (also called delta function normalizable) because
they are simply a phase.

We now recall the fact there there is a one parameter family of
solutions $u(T)$ to the differential equation~(\ref{ueq}), each
one of which corresponding to a different state $\Psi_u$.  If we
label each solution $u_w(T)$ by its value $w := u(T_0)$ at some
initial time $T = T_0$ then we may construct (truly) normalizable
wavefunctions by making ``wave packets'' of these solutions
\begin{equation}
\Psi_f(c_1, c_2, c_3; T) := \int dw \ f(w) \Psi_{u_w} = \int dw \
f(w) e^{i\frac{R^3}{\ell_p V}c_1 c_2 c_3 (1 + u_w)}
\end{equation}
where $f(w)$ is a normalizable function with support on a suitable
interval similar to that discussed in~\cite{lqginf}\footnote{In
that paper~\cite{lqginf}, it was shown that that the value of the
function $u(T)$ is constrained such that the metric remain real
and maintain a Lorentzian signature.  Hence, $w$ is similarly constrained to
such values.}. By normalizable, we mean that
\begin{equation}
\int dw \ |f(w)|^2 < \infty.
\end{equation}
This type of normalization is not possible in the Kodama
state~(\ref{kodhomo}) as there is no such one-parameter family of
solutions to take wave packets of.

Here we see that, by introducing a scalar field into our model, we
are able to describe the gravitational field (encoded by the
configuration space variables $c_I$) with respect to the scalar
field $T \approx \phi$.  In this framework, the modified Kodama
state~(\ref{psiu}), which is only continuously normalizable in the
standard inner product, is able to form a normalizable state by
making wave packets.

\section{The Spatially Flat, Isotropic Model: A Special Case of Bianchi I}
\label{sec:flatiso} Before discussing the implications of the
above results, we wish to write the solution of the previous
section specifically for a spatially flat and isotropic spacetime.
In this case, the real, homogeneous configuration space variables
are all equal
\begin{equation}
c_1 = c_2 = c_3 =: A
\end{equation}
as are their conjugate momenta
\begin{equation}
p^1 = p^2 = p^3 =: E.
\end{equation}
The Hamiltonian~(\ref{hamb}) then reads
\begin{equation}
H = 2 R^3 A^3 \left[ 3 \frac{1}{A^2} E \frac{1}{A^2} E \left(1 -
\frac{\ell_p^2 V}{3 A^2} E \right) \right]^{\frac{1}{2}}
\end{equation}
where we have written the terms in the order in which they will be
written in the corresponding Hamiltonian operator.

We will now write the classical and quantum solutions without much
commentary as the derivation is completely analogous to that of
the previous section.  The classical Hamilton-Jacobi function is
\begin{equation}
\label{isohj} S_v(A, T) = \frac{R^3}{\ell_p V(T)} A^3 (1 + v(T))
\end{equation}
where $v$ is a function of $T$ alone and satisfies the ordinary
differential equation
\begin{equation}
\dot{v} = \frac{ \dot{V}}{V}(1 + v) + \frac{6i}{\ell_p} (1 + v)
\sqrt{3 v}.
\end{equation}
The representation of the $\hat{A}$ and $\hat{E}$ operators can be
taken as
\begin{equation}
\label{isorep} \hat{A} \Psi = A \Psi, \qquad \hat{E} \Psi = -
\frac{i}{R^3 \ell_p} \frac{\partial}{\partial A} \Psi
\end{equation}
and the corresponding solution to Schr\"odinger's
equation~(\ref{schro}) is
\begin{equation}
\Psi_v := e^{i S_v} = e^{\frac{i R^3}{\ell_p V} A^3 (1+v)}
\end{equation}
which, as in~(\ref{psiu}), is also the semi-classical state.

This wavefunction $\Psi_v$ is different from the wavefunction
derived in~\cite{lqginf} using the same framework.  The difference
is due to the use of different operator orderings in the
Hamiltonian. The benefit of the
solution in this paper is that it stems from the more general
Bianchi I solution~(\ref{psiu}) whereas the solution described
in~\cite{lqginf} cannot be generalized. Furthermore, $\Psi_v$ is a
semi-classical state that is also a full solution to
Schr\"odinger's equation and, in this respect, is more closely
related to the Kodama state, as was discussed in the previous
section.  The wavefunction of~\cite{lqginf} does not share this
property but is in fact written as a \emph{modified}
semi-classical state based on the classical
solution~(\ref{isohj}).

\subsection{The Spectrum of the Scale Factor}
\label{sec:interpret}  As a simple example of how one may go about
gaining information from such a quantum cosmological framework, we
can ask what the spectrum of eigenvalues is for the operator
corresponding to the classical scale factor in a flat, isotropic
universe. Recalling that $E_i$ is just the densitized triad of the
spatial manifold, one can easily show that the metric takes the
form
\begin{equation}
ds^2 = - N^2 dT^2 + E(T) ((dx^1)^2 + (dx^2)^2 + (dx^3)^2)
\end{equation}
where $N$ is the \emph{undensitized} lapse function described in
detail in~\cite{lqginf}.  Hence, $E(T)$ is the square root of the
scale factor.

Since the operator $\hat{E}$ is simply a (scaled) derivative
operator~(\ref{isorep}), its eigenvalues are analogous to the
familiar momentum eigenvalues from ordinary quantum mechanics. The
eigenvectors of $\hat{E}$ are given by
\begin{equation}
\psi_E(A) = e^{i E R^3 \ell_p A}
\end{equation}
and are labelled by the real eigenvalues $E$. As in ordinary
quantum mechanics, if the domain of $A$ is $\mathbb{R}$ (as was
assumed in~\cite{lqginf} and in this paper), then the spectrum of
$\hat{E}$ is continuous. This runs contrary to the prediction of
loop quantum gravity~\cite{carlo} and, more specifically, to the
cosmological predictions of Martin Bojowald~\cite{bojvolop}, that
the spectrum of volume eigenvalues be
quantized.  If the scale factor had discrete eigenvalues, it would
suggest that the universe cannot evolve continuously in a
classical manner but must proceed by some quantum evolution. The
fact that we do not find this in our model may be a hint that our
approach to quantum cosmology is too naive.

Another possibility that this result may suggest is that the
domain of $A$ should not be taken as $\mathbb{R}$ but as a finite
interval. If quantum states were required to have periodic
boundary conditions on such a finite interval then the spectrum
would be quantized just as angular momentum is quantized in
ordinary quantum mechanics.  Further study is required in order to
resolve this issue.

\section{Discussion and Future Projects}
\label{sec:disc} We have shown that the
incorporation of a scalar field into the Ashtekar-Sen formalism of
general relativity allows for a choice of time gauge which leads
to a standard Hamiltonian theory.  By restricting this formalism
to a class of homogeneous spacetimes (a subclass of Bianchi class
A), we obtain a Hamiltonian theory with a finite number of degrees
of freedom that describes a homogeneous universe filled with a
scalar field. The finite number of degrees of freedom allows us to
straightforwardly quantize to obtain a corresponding quantum
framework to describe inflationary cosmology.

By looking strictly at the Bianchi I model we are able to find a
one-parameter family of classical solutions.  That is, for every
solution $u(T)$ of the differential equation~(\ref{ueq}) we obtain
a homogeneous spacetime solution given by the Hamilton-Jacobi
function $S_u$~(\ref{su}).  The solutions $u(T)$ also label a
class of quantum states~(\ref{psiu}) that are full solutions to
the corresponding quantum theory.  These quantum states are the
semi-classical states $e^{iS_u}$ corresponding to the classical
solutions given by $S_u$.  We considered flat and isotropic
spacetimes as a special case of Bianchi I in order to provide an
alternative solution to that give in~\cite{lqginf}.

We close our discussion with a  brief review of what is left to be
done and some possible directions this work may go in the future.

\bigskip

\textbf{Reality conditions and a physical inner product.}  In the
Bianchi class A framework of section~\ref{sec:classa}, the
classical configuration space variables $c_I$ were, in general,
complex numbers.  A quantization of such a complex Hamiltonian
theory requires a specially constructed inner product so that
expectation values of real quantities turn out to be real. This
problem was avoided in the Bianchi I case because the
3-dimensional spin connection vanished and we were able to
redefine the theory in terms of purely real variables.  In such a
framework, the standard inner product suffices.  For other Bianchi
models, the spin connection does not vanish identically, the $c_I$
cannot be taken to be real, and a new inner product will need to
be defined. In order to have a well-defined quantum framework for
the class A models, such an inner product will need to be found
that is applicable to all models.

\bigskip

\textbf{Hermitian ordering of the Hamiltonian.}  Throughout this
paper, and in~\cite{lqginf}, we have used an operator ordering
that leads to a non-Hermitian Hamiltonian operator. As is well
known, a Hamiltonian must be Hermitian in order for probability to
be conserved.  Hence, it may be required to solve the quantum
theory using a Hermitian ordering of the Hamiltonian in order to
obtain a probabilistic interpretation for the wavefunction
solution.  If this is, indeed, required, it is hoped that the
techniques used in this paper will lead to a similar solution to
the Schr\"odinger equation for the new Hamiltonian. Such a
solution has been searched for without any success to date.

However, given the difficulty in interpreting \emph{any}
wavefunction of the universe it is not clear that the ordinary
interpretation of quantum mechanics can be applied to a theory of
quantum cosmology.  For this reason, we have gone ahead with the
non-Hermitian Hamiltonian operator as it leads to an exact
solution analogous to the Kodama state.  Of course, such a
solution will also be a solution to the Schr\"odinger equation
using a Hermitian Hamiltonian \emph{at the semi-classical level}.

\bigskip

\textbf{Numerical analysis.} All of the cosmological solutions
based on the framework of this paper have depended on the solution
to an ordinary differential equation
(\textit{e.g.}~equation~(\ref{ueq})) which cannot be written in
closed form. Some initial numerical analysis has been done on such
equations~\cite{lqginf} but, to fully understand the dynamics of
the corresponding quantum solutions~$\Psi_u$~(\ref{psiu}), it will
be necessary to better understand the dynamics of $u(T)$.  It
seems that the best way to do this is through computational
integration of the differential equation.

Another motivation for the research discussed in this paper was to
examine the dynamics of a quantum Bianchi IX universe in which the
symmetry group of space is $SU(2)$.  The study of classical
spacetimes suggests that the dynamics of the spatial geometry of a
general manifold close to a singularity (such as the big bang
singularity) can accurately be modelled by a Bianchi IX
spacetime~\cite{bkl} and that the evolution of such a Bianchi IX
model is chaotic~\cite{ryshep, bianchaos}.  A recent result of the
loop quantum cosmology program~\cite{bojrev} is that loop quantum
gravity effects become important close to such classical
singularities and that such effects curb the chaotic evolution of
the quantum Bianchi IX model which lead to a more regular
evolution of the spatial geometry in such regimes~\cite{bojbianc1,
bojbianc2}.

It will be interesting to study the Bianchi IX model in our
framework to see whether or not the evolution is chaotic close to
classical singularities.  Since an exact wavefunction solution was
not found for the Bianchi IX model, it may be necessary to evolve
an initial state numerically using the Bianchi IX Hamiltonian
operator (the operator version of equation~(\ref{hamiltonian})
with $n^I \equiv 1$ for all $I$) in order to proceed with this
analysis.

\bigskip

\textbf{Spatial perturbations and the cosmic microwave
background.}  The greatest success of standard, inflationary
cosmology is the accurate prediction of the spectrum of
perturbations observed in the cosmic microwave background
(CMB)~\cite{liddle}.  Due to the exponential expansion of space
during the inflationary epoch and the dynamics of quantum scalar
fields in such spacetimes, scales well below the Planck length are
expanded to macroscopic scales in most inflationary
models~\cite{transplanck}.  Since many believe that low energy
physics no longer applies at such small scales, the precise
measurement of the CMB may provide the first observational
evidence for quantum gravity.  Indeed, a recent semi-classical
analysis~\cite{lqgcmb} suggests loop quantum gravitational effects
may leave an observational signature in the CMB.

The general framework described in this paper may prove to be an
advantageous starting point for a more accurate calculation of the
CMB spectrum that incorporates the quantum nature of the
gravitational field.  While very little work has been done in this
regard, one possible direction is to follow an analogous
linearization procedure to that described in~\cite{linkod} about
one of our Bianchi models. In that paper, the authours are able to
define an inner product which is necessary for any computation.
While they find that the linearized Kodama state is
non-normalizable (in the Lorentzian theory) it is hoped that the
modifications introduced by the scalar field and the choice of
time gauge described in this paper will lead to a normalizable
state. This was indeed the case for the purely homogeneous states,
as described in section~\ref{sec:normal}.  The successful
calculation of the CMB spectrum from a theory of quantum gravity
would then give us an observable prediction which could either
support or falsify our approach to quantum cosmology.

\section*{Acknowledgements}
The authour would like to thank Lee Smolin and Stephon Alexander
for their continued collaboration in this research program and for
useful discussions during the course of this work.


\begin{thebibliography}{99}

\bibitem{liddle} A.R.~Liddle and D.H.~Lyth
\textit{Cosmological Inflation and Large-Scale Structure}
Cambridge University Press: Cambridge (2000).

\bibitem{carlo} C.~Rovelli \textit{Quantum Gravity} To be published by
Cambridge University Press in October, 2004.
Available online at http://www.cpt.univ-mrs.fr/\~{}rovelli/book.pdf.

\bibitem{thiemann} T.~Thiemann \textit{Introduction to Modern
Canonical Quantum General Relativity} [arXiv:gr-qc/0110034].

\bibitem{carlorev} C.~Rovelli "Loop Quantum Gravity",
Living Rev. Relativity \textbf{1}, 1 (1998). [Online article]:
cited on \today, http://www.livingreviews.org/lrr-1998-1.

\bibitem{leerev} L.~Smolin \textit{Quantum Gravity with a Positive
Cosmological Constant} [arXiv:hep-th/0209079].

\bibitem{ashvar} A.~Ashtekar \textit{New Variables for Classical
and Quantum Gravity} Phys. Rev. Lett. \textbf{57} 2244 (1986);
\textit{New Hamiltonian Formulation of General Relativity} Phys.
Rev. \textbf{D36} 1587 (1987).

\bibitem{ashbook} A.~Ashtekar \textit{Lectures on Non-Perturbative
Canonical Gravity} World Scientific: Singapore (1991).

\bibitem{bojrev} M.~Bojowald and H.A.~Morales-T\'ecotl
\textit{Cosmological Applications of Loop Quantum Gravity} Lect.\ Notes Phys.\
\textbf{646}, 421 (2004) [arXiv:gr-qc/0306008].

\bibitem{bojsing} M.~Bojowald \textit{ABsence of a Singularity in Loop Quantum Cosmology} Phys.\ Rev.\ Lett.\ \textbf{86} 5227 (2001) [arXiv:gr-qc/0102069].

\bibitem{lqginf} S.~Alexander, J.~Malecki and L.~Smolin
\textit{Loop Quantum Gravity and Inflation}, Phys.\ Rev.\
\textbf{D70} 044025 (2004) [arXiv:hep-th/0309045].

\bibitem{kodama} H.~Kodama, Prog.~of Theo.~Phys.~\textbf{80}, 1024
(1988); Phys. Rev. D \textbf{D42} 2548 (1990).

\bibitem{dirac} P.A.M.~Dirac \textit{Lectures on Quantum
Mechanics} Belfer Graduate School of Science Monographs
\textbf{2}, Yeshiva University Press: New York (1964).

\bibitem{ryshep} M.P.~Ryan and L.C.~Shepley \textit{Homogeneous
Relativistic Cosmologies} Princeton University Press: Princeton
(1975).

\bibitem{kodbianchi} H.~Kodama \textit{Specialization of
Ashtekar's Formalism to Bianchi Cosmology}, Progr.~of
Theo.~Phys.~\textbf{80}, 6, 1024 (1988).

\bibitem{homobojo} M.~Bojowald \textit{Homogeneous Loop Quantum
Cosmology} Class. Quant. Grav. \textbf{20} 2595 (2003)
[arXiv:gr-qc/0303073].

\bibitem{bojbianc1} M.~Bojowald, G.~Date and K.~Vandersloot
\textit{Homogeneous Loop Quantum Cosmology: The Role of the Spin
Connection} Class. Quant. Grav. \textbf{21} 1253 (2004)
[arXiv:gr-qc/0311004].

\bibitem{diracprinc} P.A.M.~Dirac \textit{The Principles of
Quantum Mechanics, 4th Ed.} Oxford University Press: Oxford
(1958).

\bibitem{goldstein} H.~Goldstein, C.~Poole and J.~Safko
\textit{Classical Mechanics} Addison Wesley: San Fransisco (2002).

\bibitem{soo} C.~Soo \textit{Wave Function of the Universe and
Chern-Simons Perturbation Theory} Class. Quant. Grav. \textbf{19}
1051 (2002) [arXiv:gr-qc/0109046].

\bibitem{bojvolop} M.~Bojowald \textit{Loop Quantum Cosmology: II.
Volume Operator} Class. Quant. Grav. \textbf{17} 1509 (2000)
[arXiv:gr-qc/9910104].

\bibitem{bkl} V.A.~Belinskii, I.M.~Khalatnikov and E.M.~Lifshitz
\textit{A General Solution of the Einstein Equations with a Time
Singularity} Adv. in Phys. \textbf{31}, 6, 639 (1982).

\bibitem{bianchaos} T.~Damour, H.~Henneaux and H.~Nicolai
\textit{Cosmological Billiards}, Class. Quant. Grav. \textbf{20},
R145 (2003) [arXiv:hep-th/0212256].

\bibitem{bojbianc2} M.~Bojowald and G.~Date \textit{A Non-Chaotic
Quantum Bianchi IX Universe and the Quantum Structure of Classical
Singularities} Phys. Rev. Lett. \textbf{92} 071302 (2004)
[arXiv:gr-qc/0311003].

\bibitem{transplanck} J.~Martin and R.~Brandenberger \textit{Trans-Planckian Problem of Inflationary
Cosmology} Phys. Rev. D \textbf{63} 123501 (2001)
[arXiv:hep-th/0005432]; \textit{Dependence of the Spectra of
Fluctuations in Inflationary Cosmology on Trans-Planckian Physics}
Phys. Rev. D \textbf{68} 063513 (2003) [arXiv:hep-th/0305161].

\bibitem{lqgcmb} S.~Tsujikawa, P.~Singh and R.~Maartens
\textit{Loop Quantum Gravity Effects on Inflation and the CMB}
[arXiv:astro-ph/0311015].

\bibitem{linkod} L.~Friedel and L.~Smolin \textit{The
Linearization of the Kodama State} [arXiv:hep-th/0310224].

\end{thebibliography}
\end{document}